\documentclass[prl,twocolumn,superscriptaddress,longbibliography]{revtex4-2}

\usepackage{amssymb,amsmath,amsfonts}
\usepackage{graphicx}
\usepackage{color,xcolor}
\usepackage{hyperref}
\usepackage{bm}
\usepackage{multirow}
\usepackage{natbib}
\usepackage{comment}

\begin{document}

\title{
Structural Transitions and Melting of 2D Ion Crystals in Traps 
}

\author{B. V. Pashinsky}

\affiliation{Department of Physics, University of Washington, Seattle, Washington 98195, USA}

\author{ A. Kato}

\affiliation{IonQ, Inc., College Park, MD, United States of America}

\author{B. Blinov}

\affiliation{Department of Physics, University of Washington, Seattle, Washington 98195, USA}

\date{\today}

\begin{abstract}
We investigate the structural properties and melting behavior of two-dimensional ion crystals in an RF trap, focusing on the effects of ion temperature and trap potential symmetry. We identify distinct crystal structures that form under varying trapping conditions and temperatures through experimental and theoretical analyses. As the temperature increases or the trap potential becomes more symmetric, we observe a transition from a lattice arrangement to elongated ring-like formations aligned along the trap axes. Our experimental and theoretical efforts enhance our understanding of phase transitions in low-dimensional, confined systems, offering insights into the controlled formation of quantum crystals for applications in quantum simulations and many-body physics.
\end{abstract}

\maketitle

\section{Introduction}
Trapped ion crystals provide a well-controlled platform for studying structural transitions and melting in low-dimensional systems. These systems, where ions self-organize under the influence of long-range Coulomb interactions and external trapping potentials, have been extensively used to explore fundamental aspects of phase transitions, many-body physics, and quantum simulations~\cite{Leibfried, Haffner}. In particular, two-dimensional (2D) ion crystals confined in radiofrequency (RF) traps offer a unique setting where the interplay between trapping anisotropy and thermal effects leads to rich structural dynamics~\cite{Diedrich, Schiffer}.  

\begin{figure}
%\isPreprints{\centering}{} % Only used for preprints
\includegraphics[width=0.45\textwidth]{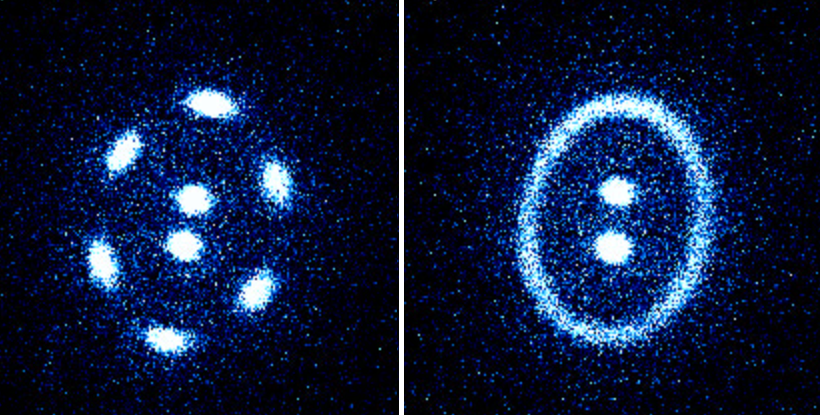}
\caption{An example of melting transition in a crystal or 8 trapped Ba$^+$ ions confined in a 2D trap as the ion temperature increases. Left: the ions form a crystalline structure with two distinct shells at a lower temperature. Right: the 6 ions of the outer shell have melted at a higher temperature. The temperature is controlled by tuning the Doppler-cooling lasers frequencies. \label{fig:experiment3}}
\end{figure}

At low temperatures, ions in a 2D trap typically arrange into stable lattice-like structures resembling the Wigner crystal~\cite{Wigner}. However, as the temperature increases, thermal fluctuations drive the system through a melting transition, where ions transition from localized vibrational motion around equilibrium positions to diffusive motion~\cite{OrMelting}, as illustrated in Fig.~\ref{fig:experiment3}. Understanding the precise nature of this transition is critical for applications in precision metrology~\cite{Keller}, quantum computing~\cite{Monroe}, and studies of non-equilibrium statistical mechanics~\cite{Schaetz}.  

In addition to temperature, the symmetry of the trap potential plays a crucial role in determining crystal structure and crystal melting. When the trapping potential is highly anisotropic, the ions form elongated, elliptical configurations, while more symmetric traps favor compact, circular arrangements. The dependence of melting behavior on trap anisotropy remains an open question, particularly regarding how structural deformations affect the critical temperature \( T_c \) and whether standard melting criteria, such as Lindemann’s criterion~\cite{Lindemann}, remain valid in this confined geometry.  

\unskip

In this work, we experimentally observe the structural properties and melting behavior of 2D ion crystals in an RF trap and investigate them using a combination of theoretical and numerical approaches. We analyze the transition from lattice-like configurations to ring-like structures as a function of temperature and trap anisotropy. The critical temperature for melting is determined using both the Lindemann criterion and free energy calculations, allowing us to assess their consistency and limitations. Furthermore, we compute the diffusion coefficient of ions as a function of temperature and compare its behavior with the melting threshold obtained from structural fluctuations. Our results provide new insights into phase transitions in low-dimensional confined systems and have implications for optimizing ion-crystal-based platforms for quantum simulations and many-body experiments.

\section{Experimental Techniques}

The experiments focus on the generation, control, and characterization of two-dimensional (2D) Coulomb crystals of trapped barium ions (\( \text{Ba}^+ \)) in a radiofrequency (RF) trap.  These ion crystals are of interest for quantum computing, quantum simulations, and fundamental studies of phase transitions in low-dimensional systems~\cite{Duan2024, Richerme2021}.

\begin{figure}
%\isPreprints{\centering}{} % Only used for preprints
\includegraphics[width=0.5\textwidth]{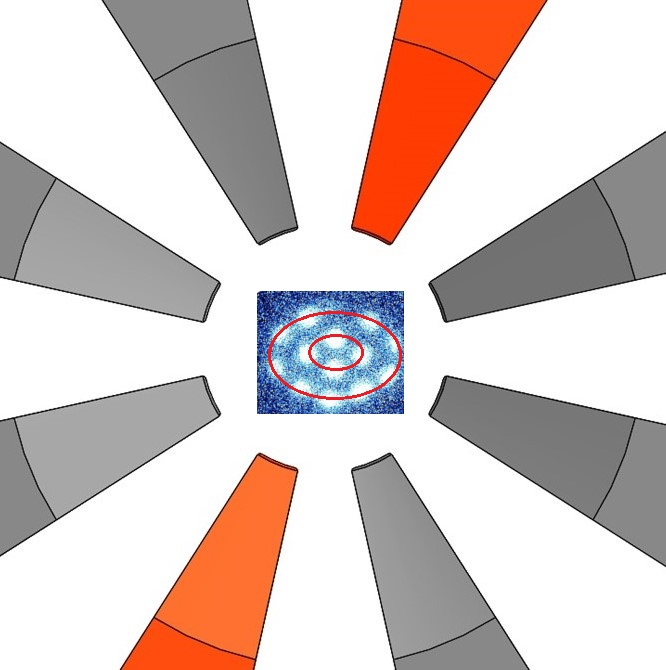}
\caption{The schematic of the top view of the ion trap used in the experiments. The eight sectors of the ring electrode are shown, with two electrodes highlighted in red that were used to control the anisotropy of the trapping potential. An image of a 14-ion 2D crystal is superimposed (not to scale), with the red lines outlining the shell structure of the crystal. \label{fig:experiment1}}
\end{figure}

The ion trap used in these experiments is a modification of the standard Paul trap, where the ring electrode is flattened and divided into eight sectors~\cite{ivory2020paul}. A diagram of the trap is shown in Fig.~\ref{fig:experiment1}. This configuration allows for independent control of the trapping potential in the plane of the crystal. Two endcap electrodes, shaped as truncated cones (not shown in the figure for clarity), ensure strong transverse confinement and provide optical access for imaging and laser addressing. The trap operates at an RF frequency of approximately 10 MHz, with the amplitude in the range of 800–1500 V applied to the endcap electrodes via a double-coil helical resonator. The resulting pseudopotential supports the formation of 2D Coulomb crystals by balancing the in-plane repulsion from ion-ion interactions and the transverse confinement from the RF field~\cite{yoshimura2015creation}. For more details about this ion trap please see~\cite{ivory2020paul}.

\unskip

Barium ions are loaded into the trap using a two-step photoionization process with a 791 nm external cavity diode laser (ECDL) and a 337 nm nitrogen laser. Once ionized, the ions are Doppler-cooled using 493 nm and 650 nm laser beams, which address the \( 6S_{1/2} \rightarrow 6P_{1/2} \) and \( 5D_{3/2} \rightarrow 6P_{1/2} \) transitions, respectively. The fluorescence of the ions is imaged using an electron-multiplying charge-coupled device (EMCCD) camera. The magnification of the imaging system is approximately 80. The laser cooling process leads to crystallization, where the ions form a radial 2D structure due to their mutual Coulomb repulsion and the symmetry of the trapping potential~\cite{kato2023pulsed}. 

A key challenge in 2D Coulomb crystals is excess micromotion, which results from the RF field and leads to time- and position-dependent Doppler shifts. Since the micromotion amplitude increases with radial distance from the trap center, ions in different regions of the crystal experience different effective cooling rates. To address this issue, augmented Doppler cooling schemes ~\cite{kato2022doppler, kato2023pulsed, Richerme2021} can be used to stabilize crystals of ~50 ions, or alternate trap geometries that minimize or eliminate the micromotion in the direction of Doppler cooling. In the latter configuration stable manipulation of large 2D crystals of over 100 ions for quantum simulation has been achieved~\cite{Kiesenhofer2023,Duan2024}.

The trapped ions form concentric shells as pictured in Fig.~\ref{fig:experiment1}, with each shell corresponding to a different radial equilibrium position. The number of shells and their spacing are determined by the interplay between the Coulomb repulsion and the trapping potential. Experimental images confirm the expected shell structure, with ion positions matching theoretical predictions. The shape of the crystal can be dynamically tuned by adjusting the DC biases on the sector electrodes or modifying the endcap voltages, allowing for transitions between nearly circular and elliptical crystal geometries.

We use this experimental data to extract information about the ellipticity, as well as the melting behavior of trapped ion crystals. During melting, the ions begin moving along the elliptical paths with increasing speed, eventually forming continuous rings in experimental images, as illustrated in Fig.~\ref{fig:experiment3}. Furthermore, the stability of the crystalline structure at different temperatures allows us to determine the threshold conditions for the transition between solid and melted states.

\section{Theoretical Approach}
We consider a system of $N$ identical charged particles of mass $m$ and charge $e$ confined in an anisotropic quadratic potential trap. The effective confinement is created by the  combination of the pseudopotential due to the high frequency RF oscillating electric field and a static electric field~\cite{Wineland1998, Leibfried2003}. In such traps, the rapid oscillations of the RF field average out, leading to a time-averaged potential that effectively confines the ions in a plane perpendicular to the one of the principal axes of the trap (taken to be the x-axis in the following discussion). The interplay between the Coulomb repulsion and the external confinement results in a self-organized ion crystal, where the equilibrium positions of the ions are determined by a balance of these competing forces.

The Hamiltonian governing the system is given by:
\begin{align}
H = \sum_{i=1}^{N} \left( \frac{p_{i}^{2}}{2m} + \frac{m}{2} \left(\omega_{y}^{2} y_{i}^{2} + \omega_{z}^{2} z_{i}^{2} \right) + \sum_{j>i}^{N} \frac{\alpha}{\left| \boldsymbol{r}_{i} - \boldsymbol{r}_{j} \right|} \right),
\end{align}
where the summation is over all the ions whose positions in y-z plane are given by $y_i$, $z_i$, and whose momenta are $p_i$, respectively, $\omega_y$ and $\omega_z$ are the secular frequencies in the plane of the crystal, and $\alpha = {e^{2}}/{4\pi\epsilon_{0}}$
is the Coulomb interaction coefficient. The secular frequencies define the degree of anisotropy $\gamma = \omega_z / \omega_y$ in the potential well, which directly affects the shape and stability of the ion crystal. %For a highly anisotropic trap, the equilibrium positions of the ions form an elongated ellipsoidal shape, while a more symmetric trap leads to a nearly circular Wigner-like crystal. 

\subsection{Dimensionless Representation of Positions}

To simplify the mathematical description, we introduce the characteristic length scale of the system using the Bohr radius of the ion:
\begin{align}
a_{0} = \frac{4\pi\epsilon_{0} \hbar^{2}}{m e^{2}},
\end{align}
where $\epsilon_{0}$ is the vacuum permittivity and $\hbar$ is the reduced Planck's constant. 
This defines a natural unit of length associated with the quantum mechanical Coulomb problem. Then, using $a_0$, we introduce the dimensionless coordinates:
\begin{align}
\boldsymbol{r} \rightarrow \left( \frac{\hbar^{2}}{m^{2} \omega_{y}^{2} a_{0}} \right)^{1/3} \boldsymbol{r}.
\end{align}
This rescaling allows us to express all quantities in terms of dimensionless parameters that highlight the fundamental interplay between the kinetic, potential, and interaction energies.

Then, the Hamiltonian of the system, in units of $ \hbar \left(\frac{\hbar\omega_{y}^{2}}{m a_{0}^{2}}\right)^{1/3} $, can written as:
\begin{align}
\label{h_less}
H_{\text{d}} = \sum_{i=1}^{N} \left( \frac{1}{2} \left( y_{i}^{2} + \gamma^2 z_{i}^{2} \right) + \sum_{j>i}^{N} \frac{1}{\left| \boldsymbol{r}_{i} - \boldsymbol{r}_{j} \right|} \right).
\end{align}
In this form, the dimensionless Hamiltonian explicitly shows how the system’s behavior is controlled by the ratio of trapping frequencies, $\gamma=\omega_z / \omega_y$, and the number of ions $N$. By varying these parameters, we can explore different structural regimes, from strongly confined Wigner crystals to fluid-like configurations at higher temperatures.

\subsection{Melting Condition}

For the finite systems, the concept of critical temperature is inherently ambiguous, as phase transitions are not sharply defined in such cases \cite{Guardiola2011, Robbins1990}. Nevertheless, structural changes occur, making temperature a key parameter in characterizing the system’s behavior. To describe this, we consider the partition function in dimensionless units of position $r$ and momentum $p$:  

\begin{equation}  
    Z \propto \int d^{2N}p \, d^{2N}r \, e^{-(T_0/T) H_{\text{d}}},  
\end{equation}  
where $k_B$ is the Botzmann constant, $T$ is the temperature and $T_0=\frac{\hbar}{k_{B}} \left(\frac{\omega_{y}^{2} \hbar}{m a_{0}^{2}}\right)^{1/3}$. 
%This formula shows that the partition function, and thus all thermodynamic properties, depend on the ratio $T_0 / T$ along with the dimensionless Hamiltonian. 
Since the Hamiltonian in Eq.~\eqref{h_less} depends only on the ion positions and the anisotropy parameter $\gamma$, we expect the critical temperature to take the general form:  

\begin{align}  
    T_c = T_0 g(\gamma, N),  
\end{align}  
where function $g(\gamma, N)$ encapsulates the effects of anisotropy and the finite number of ions. This scaling implies that, when expressed in units of $T_0$, the melting temperature should primarily depend on $\gamma$ and $N$, enabling direct comparisons between theory and simulations.  

Additionally, due to the system’s rotational symmetry, we have the property  $g\left( \gamma,  N\right) = g\left( \gamma^{-1},  N\right) $  which is particularly useful, as it allows us to analyze the system’s behavior by focusing only on cases where $\gamma \leq 1$.  

Using Lindemann’s criterion, which states that a system remains crystalline as long as the amplitude of atomic vibrations does not exceed a fraction \(c\) of the interparticle spacing \cite{Lindemann, Zheng1998, Zhou1991}, we can analyze the stability of the crystal structure. Instead of calculating the amplitudes directly, we can compare the scales of potential energy \(V \sim k_B T_0\) and kinetic energy \(K \sim k_B T\). Since energy depends quadratically on the amplitude, the critical temperature can be estimated as:
\begin{align}
T_{c} = c^2 T_0 = c^2 \frac{\hbar}{k_{B}} \left(\frac{\omega_{y}^{2} \hbar}{m a_{0}^{2}}\right)^{1/3}.
\end{align}
For typical ion trap frequencies ($\omega_{y,z} \sim 2\pi\times100$ kHz), we obtain $T \sim 0.01 K$. This corresponds to an approximation where the function $g(\gamma,N)$ is treated as constant $c^2$. This simple estimation provides a physically meaningful order of magnitude. However, it does not account for the dependence on the number of ions or the anisotropy parameter. In the following subsections, we develop a more precise approach based on Lindemann’s criterion, incorporating a more detailed analysis of structural fluctuations.

\subsection{Toy-Model Theory of Melting}
At low temperatures, ions in a 2D trap arrange into a series of concentric elliptical shells due to the anisotropic confinement. The equilibrium positions of ions $(y_{i},z_{i})$ within a given shell satisfy the constraint:

\begin{align}
y_{i}^{2} + \gamma^2 z_{i}^{2} = R_{n}^{2},
\end{align}
where $R_n$ is the characteristic radius of the $n$-th shell. 
%The elliptical shape results from the anisotropy of the trapping potential, which distorts the otherwise circular symmetry of the Coulomb crystal.

A key idea of this toy model is that an ion in the outermost shell is effectively confined within a potential well created by the Coulomb interaction with the inner shell, which consists of $N_n$ ions. Experimental observations indicate that melting predominantly occurs via angular motion rather than radial motion, suggesting that the angular coordinate $\varphi$ is the "soft" degree of freedom in the system. In contrast, the radial coordinate is effectively constrained by the ellipticity of the trap, meaning that an ion radius is dictated by the equation of the ellipse. As a result, the system's dynamics can be effectively described using only the angular coordinate, reducing the problem to a one-dimensional potential landscape along $\varphi$.  

Furthermore, we assume that the outermost shell melts before the inner ones, which is supported by experimental observations. This implies that the angular displacements become more pronounced closer to the boundary of the ion crystal. Consequently, in our model, we consider the inner ions to be nearly immobile compared to those in the outermost shell, as their motion is significantly restricted by the stronger Coulomb forces and the trap potential. This approximation simplifies the description of the melting process while retaining the essential features observed in the experiment.

Another important aspect of this model is the role of the trapping potential. While the external confinement plays a crucial role in shaping the overall crystal structure, its effect on individual ions within the same shell is approximately uniform. This is due to the fact that, for a given shell, all ions share the same elliptical constraint. Thus, their trapping potential energy remains nearly constant, and the variations in the total energy are dominated by the Coulomb interaction with the inner shell ions.  

To estimate the effective potential barrier that an outer-shell ion must overcome to delocalize, we model the interaction energy of an ion at angular position $\varphi$ and radial position $r$ due to all $N_n$ ions in the lower shell as:
\begin{align}
V(\varphi) = \sum_{i=1}^{N_{n}} \frac{1}{\sqrt{r_{i}^{2} + r^{2} - 2r_{i} r \cos\left(\theta_{i} - \varphi\right)}}.
\label{potential}
\end{align}
Here, $r_i $ and  $\theta_{i}$  are the radial and angular coordinates of the $i$-th ion in the inner shell, which satisfy the condition:

\begin{align}
   r_{i}^{2} = \frac{2 R_{n}^{2}}{1 + \gamma^2 + \left(1-\gamma^2\right) cos\left(2\theta_{i}\right)},
\end{align}
while the radial position $r$ of the outer-shell ion satisfies:
   \begin{align}
   r^{2} = \frac{2R_{n+1}^{2}}{1+\gamma^2+\left(1-\gamma^2\right)\cos\left(2\varphi\right)}.
   \end{align}
These expressions account for the elliptical distortion of the shells due to the anisotropic trapping potential.

To further simplify this toy model while retaining the essential physics, we assume that the ions within a given shell are distributed uniformly, such that:
   \begin{align}
   \theta_{i} = \frac{2\pi i}{N_{n}}, \quad i = 1, \dots, N_n.
   \end{align}
This approximation is precise only for zero anisotropy, but it remains a good approximation even at modest non-zero anisotropy as well.

\begin{figure}
%\isPreprints{\centering}{} % Only used for preprints
\includegraphics[width=0.5\textwidth]{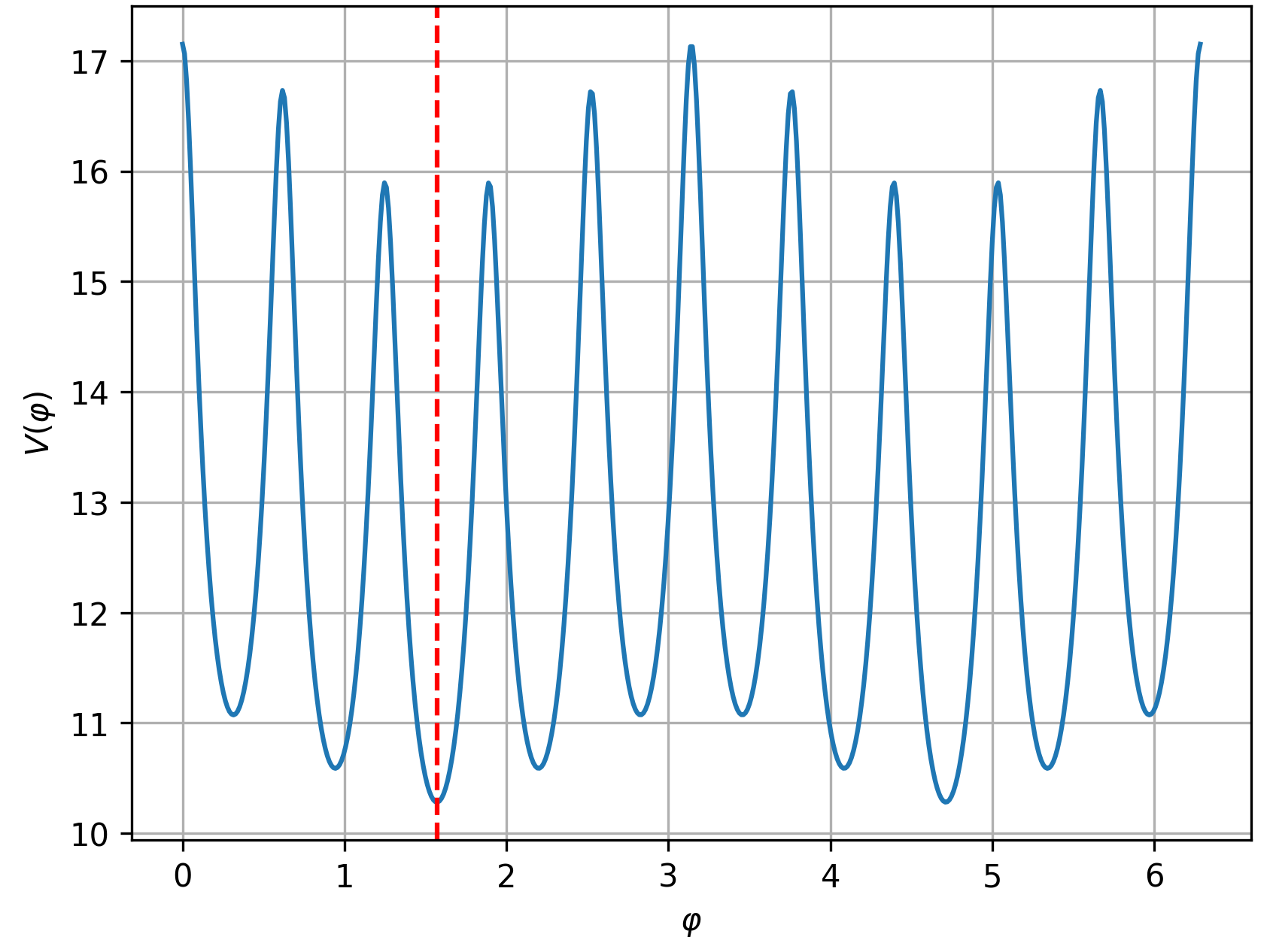}
\caption{Dimensionless potential energy $V(\varphi)$ of the outer shell ion as a function of its angular position for the trap anisotropy $\gamma = 0.9$. The red dashed line indicates one of the global minima.\label{fig:potential}}
\end{figure}

With these assumptions, the calculated total potential energy of an outer-shell ion as a function of its angular coordinate $\varphi$ is plotted in Fig.~\ref{fig:potential} with 10 ions in the shell, for the anisotropy of $\gamma = 0.9$. This function exhibits two notable features: global minima at 
$ \varphi_{min} = \frac{\pi}{N_{n}} \left( 2\left\lfloor \frac{N_{n}-2}{4} \right\rfloor +1 \right) $, and global maxima at $ \varphi_{max} = 0, \pi $, representing the highest energy barrier an ion must overcome to become delocalized within the shell. Here, the square brackets $\lfloor \cdot \rfloor$ denote the floor function, which rounds down to the nearest integer.

\unskip

Since melting occurs when the thermal energy $k_B T$ becomes comparable to the depth of this potential well, the critical temperature can be estimated by evaluating the potential energy difference between these two points:

\begin{align}
\label{toy-temperature}
\frac{T_{c}(\gamma)}{T_0} = V(\varphi_{max}) - V \left(\varphi_{min}  \right).
\end{align}

\begin{figure}
%\isPreprints{\centering}{} % Only used for preprints
\includegraphics[width=0.5\textwidth]{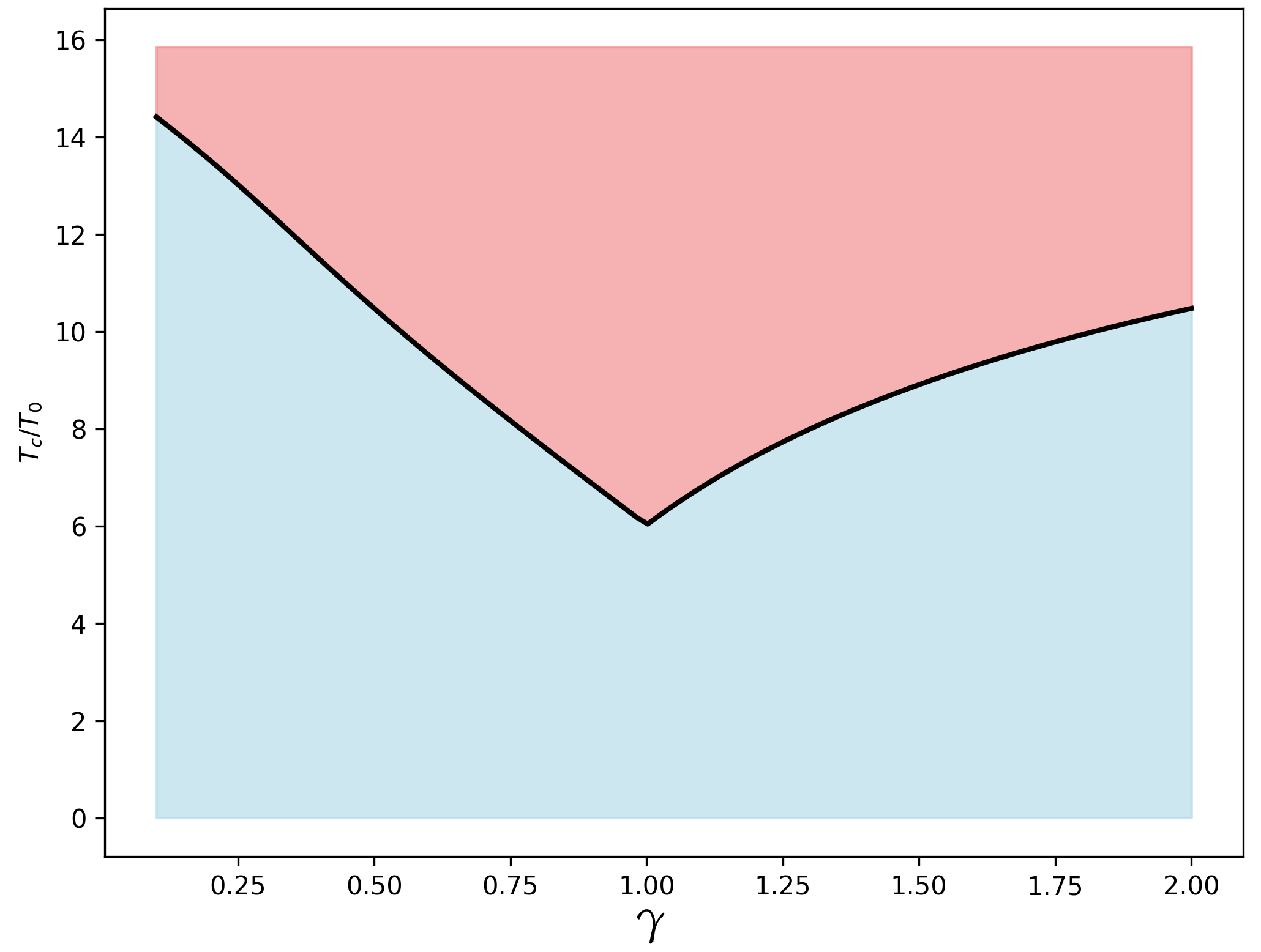}
\caption{Toy-model phase diagram. The critical temperature $T_c$ in units of $T_0$ is plotted against the anisotropy parameter $ \gamma$. The light-blue shaded region corresponds to the crystalline state, while the pink-shaded regions corresponds to the melted state. The minimum of the critical temperature is a the point of $\gamma=1$, as expected.
\label{fig:theory}}
\end{figure}

Using this equation, we can construct the theoretical phase diagram shown in Fig.~\ref{fig:theory}, which illustrates how the estimated critical temperature depends on the anisotropy parameter $ \gamma$.

\unskip

\section{Numerical Simulations}

In our study, we focus on systems with 14 and 28 ions because these configurations allow for a well-defined analysis of melting, as all ions form distinct shells. For other numbers of ions, equilibrium structures can exhibit significant variations, with some ions not fitting into well-formed shells. This structural variability complicates the definition of melting since different equilibrium states may correspond to different probability distributions for melting as a function of temperature and anisotropy. Moreover, the formation of a specific ion configuration in a 2D Coulomb crystal is not deterministic in these conditions, even when trapping parameters and ion numbers remain the same. Due to the presence of multiple energetically similar configurations, it is challenging to consistently reproduce an identical crystal structure across different experimental realizations. This intrinsic uncertainty in equilibrium states contributes to variations in melting behavior and critical temperature estimations. Similar challenges in achieving reproducible crystal configurations have been observed experimentally, as discussed in Ref.~\cite{Kiesenhofer2023}, where structural differences arising from small perturbations in initial conditions were analyzed. Understanding this variability is crucial for improving control over trapped ion systems and refining theoretical models of their phase transitions.

To analyze the melting transition of trapped ion crystals, we employ molecular dynamics simulations. This approach allows us to investigate the system’s behavior at different temperatures and extract key thermodynamic properties, such as the Lindemann criterion for melting and the probability of phase transition. The melting probability is determined by the fraction of simulations in which the crystalline structure is lost at a given temperature and anisotropy parameter.

\subsection{Molecular Dynamics Simulations}

The molecular dynamics simulations are based on integrating the classical equations of motion:
\begin{align}
    m_i \ddot{\boldsymbol{r}}_i = -\nabla_i V,
\end{align}
where $V$ is the total potential energy, consisting of the harmonic trapping potential and the Coulomb interaction between ions.

To numerically integrate the equations of motion, we use the Verlet algorithm \cite{Verlet1967}, which ensures energy conservation and stability over long simulation times. The time step is chosen to be sufficiently small to resolve the characteristic vibrational frequencies $\omega_y$ and $\omega_z$.

Unlike traditional thermostats such as Langevin dynamics \cite{Berendsen1984}, we set the temperature through the initial kinetic energy of the ions:
\begin{align}
    K = N k_B T.
\end{align}
The initial positions of the ions are set at the calculated equilibria, corresponding to the minimum of potential energy. This configuration ensures that the system starts in the crystalline state at low temperatures. Such process is much faster than the Langevin dynamics in which one has to generate random numbers on every time step, but provides identical results.

To determine whether the system undergoes melting, we apply Lindemann’s criterion for each ion. If the displacement exceeds the threshold for a given ion, that ion is considered to have "melted."

\begin{figure}[h]
%\isPreprints{\centering}{} % Only used for preprints
\includegraphics[width=0.5\textwidth]{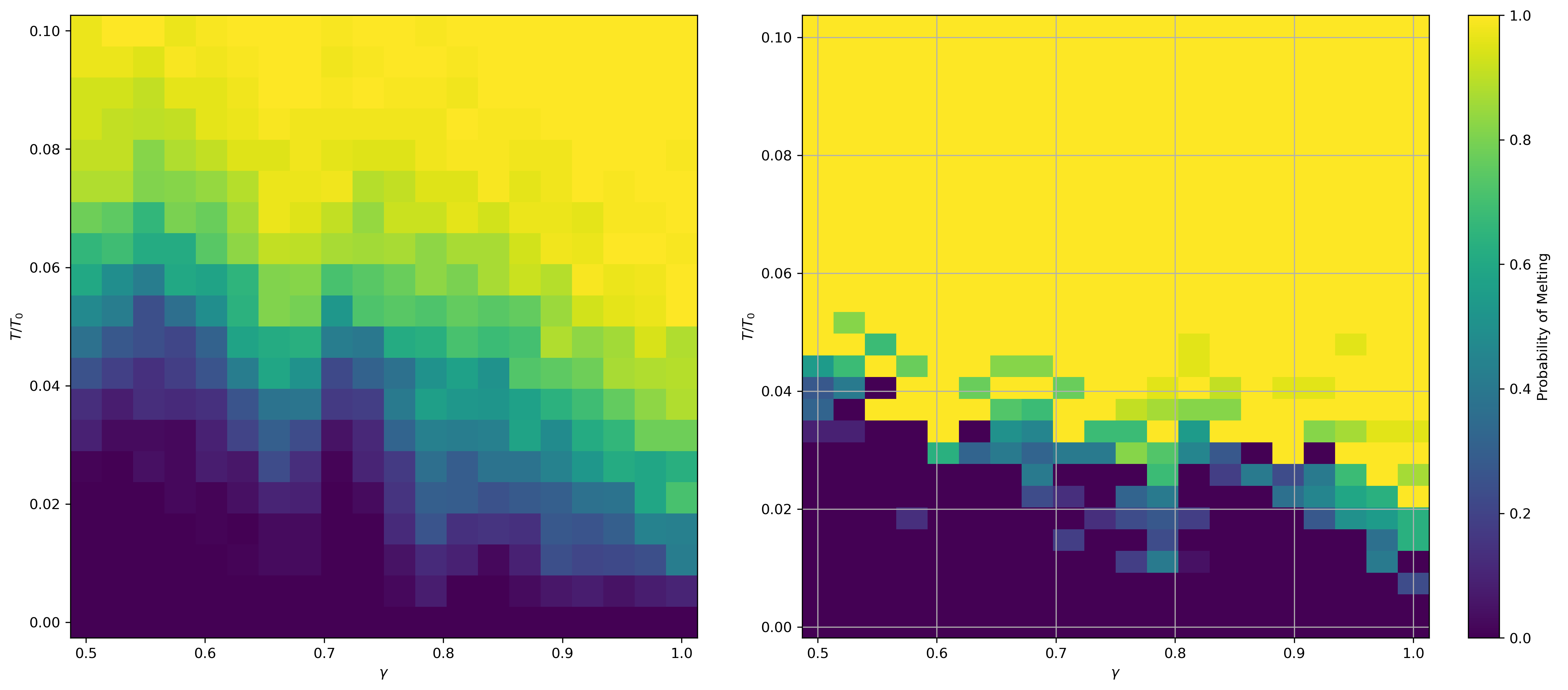}
\caption{Melting probability for different temperatures and anisotropies for crystals contained 14 ions (left) and 28 ions (right). Fewer numerical simulation runs were done for the latter, resulting is a noisier results. Nevertheless, the lower overall melting temperature for the larger crystal is evident. \label{fig:MD}}
\end{figure}

The melting probability is then estimated as the fraction of simulations in which at least one ion meets this criterion at a given temperature and anisotropy parameter. This probability is statistically averaged over multiple simulation runs to obtain a reliable phase transition data which is plotted in Fig.~\ref{fig:MD}. For 14 ions, we performed 400 simulation runs for each pair of $T$ and $\gamma$,  whereas for 28 ions, we only did 25 runs due to the increased computational complexity and limited computational power.

\unskip

\subsection{Diffusion Coefficient Estimation}

The transition to the liquid phase is signaled by a sudden increase in diffusion coefficient $D$, marking the melting temperature. This approach has been successfully applied to analyze phase transitions in trapped ion systems \cite{Wineland1998, Leibfried2003}. Similar methodologies have been employed in molecular dynamics studies of Coulomb crystals \cite{Dubin1999}. To quantify the melting transition, we compute the mean squared displacement (MSD) for all ions:
\begin{align}
\text{MSD}(t) = \frac{1}{N} \sum_{i=1}^{N} \langle |\boldsymbol{r}_i (t) - \boldsymbol{r}_i (0) |^2 \rangle.
\end{align}
For the crystalline phase, the ion motion remains bounded, and the MSD saturates at a finite value, while in a liquid-like state, ions exhibit diffusive motion, leading to a linear increase of MSD at long times. The diffusion coefficient is extracted from this long-time behavior as:
\begin{align}
D = \lim_{t \to \infty} \frac{\text{MSD}(t)}{4t}.
\end{align}

\begin{figure}[h]
%\isPreprints{\centering}{} % Only used for preprints
\includegraphics[width=0.5\textwidth]{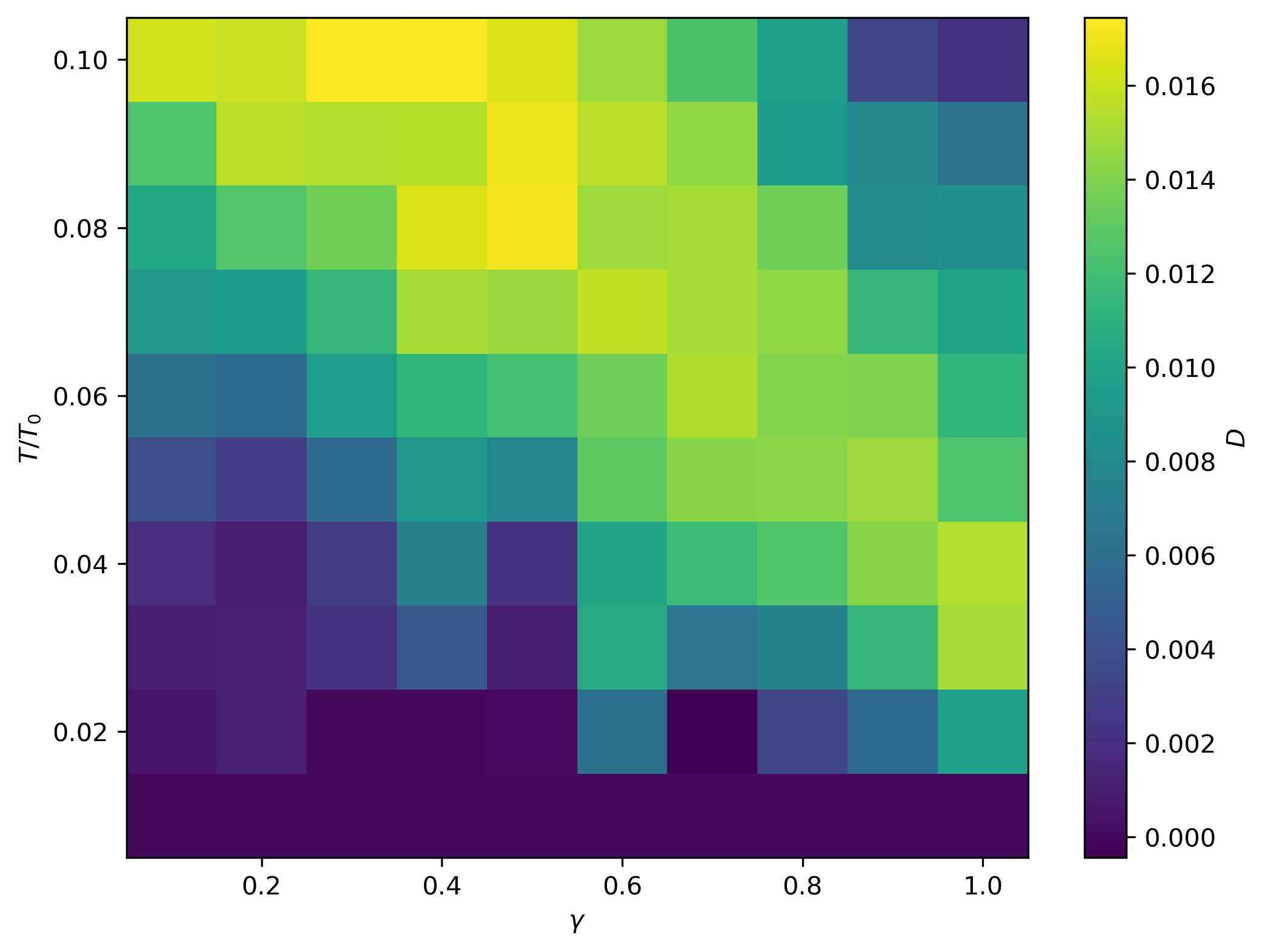}
\caption{Diffusion coefficient for different temperatures and anisotropies for the 14-ion crystal.\label{fig:difusion}}
\end{figure}

By evaluating $D$ as a function of temperature $T$ and anisotropy parameter $\gamma$, we obtain the diffusion behavior of the system, as shown in Fig.~\ref{fig:difusion}. While the diffusion coefficient provides an independent measure of melting, it is important to note that extracting $T_c$ from $D(T)$ requires computing numerical derivatives, which introduces additional sources of error due to finite-difference approximations.

Additionally, the diffusion-based approach is computationally more expensive compared to the direct probability-based method. Calculating the diffusion coefficient requires long-time averaging for accurate estimation, thus the resulting phase boundary has a larger error. At the current level of statistical uncertainty, with a maximum error of 0.00334, the melting temperatures derived from the diffusion analysis and those obtained from the melting probability method may either coincide for different anisotropies or exhibit significant discrepancies.

\unskip

There are regions where the diffusion coefficient decreases with increasing temperature (for example for $\gamma=1$), which may seem counterintuitive since higher temperatures generally lead to greater disorder and increased ion mobility. However, because melting in this system occurs ring by ring, ions are not free to move arbitrarily but are instead constrained to motion along their respective rings. At certain temperatures, this constraint may effectively reduce diffusion, as ions experience increased interactions with their neighbors while still being partially confined. Additionally, structural rearrangements within the rings could temporarily stabilize ion positions before full melting occurs, further suppressing diffusion. This behavior suggests that melting is not a simple monotonic process but rather involves intermediate structural transitions where diffusion can momentarily decrease.

\section{Discussion}
Despite the relatively small size of the simulated system, our results successfully capture the phase transition behavior of two-dimensional ion crystals in an RF trap. By systematically analyzing the melting process, we have obtained phase diagrams for systems with 14 and 28 ions. A key observation is that the critical temperature for melting is consistently lower for the 28-ion system across all values of anisotropy, as shown in Fig. \ref{fig:MD}. As seen in Eqs. \eqref{toy-temperature} and \eqref{potential}, the critical temperature decreases with increasing $R_n$ and $R_{n+1}$. Since the radii of the outermost and second-to-last rings grow as the number of ions in the system increases, the overall effect leads to a reduction in the critical temperature for larger ion crystals. This aligns with the observed trend that increasing system size results in a lower melting threshold, as the spatial expansion of the outermost rings weakens the effective confinement and reduces the characteristic energy scale of structural fluctuations.

\begin{figure}[h]
%\isPreprints{\centering}{} % Only used for preprints
\includegraphics[width=0.5\textwidth]{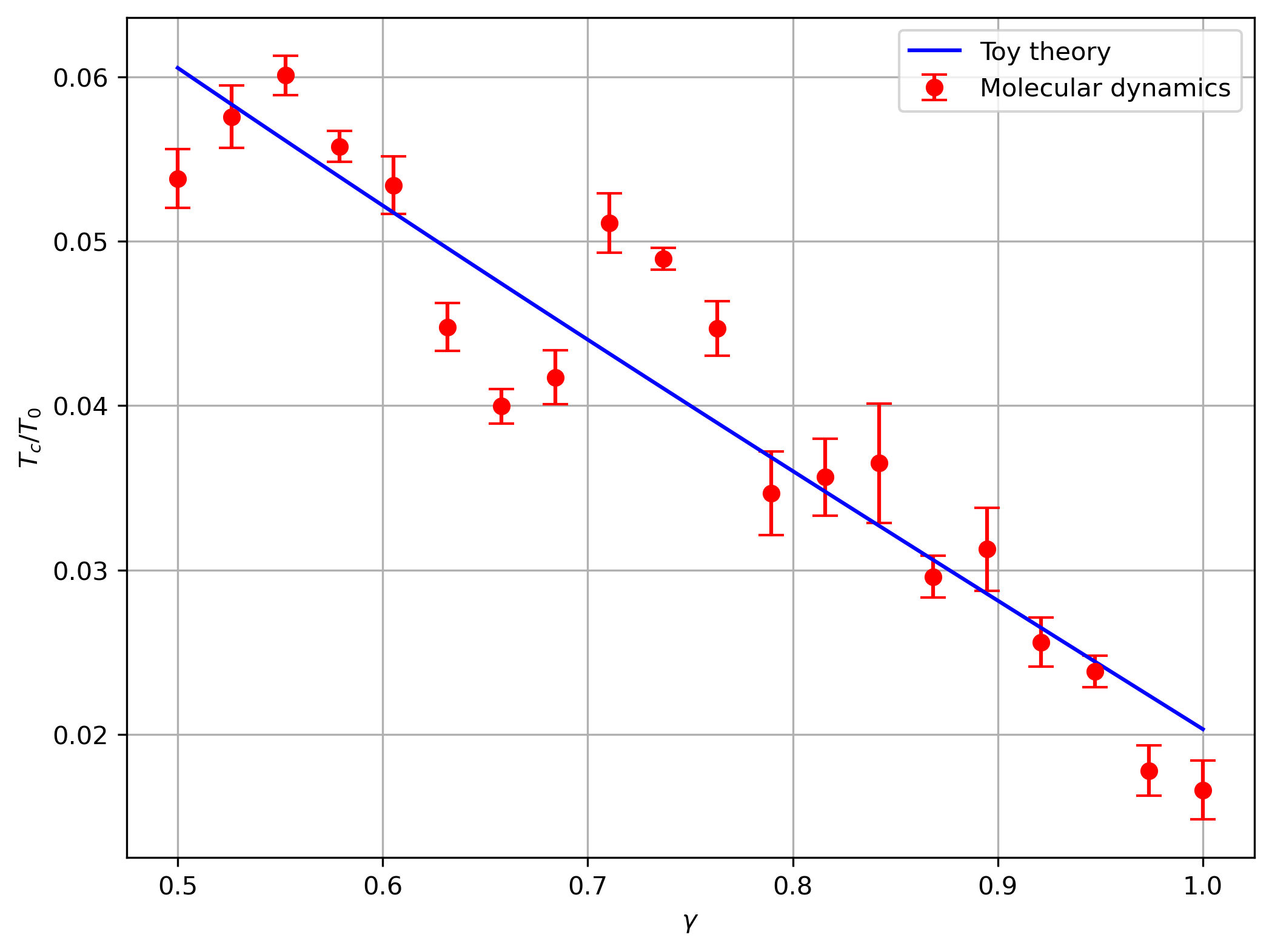}
\caption{Comparison of the critical temperature $T_c$ as a function of the crystal anisotropy $\gamma$ between toy-model theory predictions and the molecular dynamics simulations for the crystal contained 14 ions. While there is general agreement, the toy model does not predict the structure observed in the simulations.  
\label{fig:theory_vs_MD}}
\end{figure}

Our theoretical model also predicts an increase in the critical temperature with increasing anisotropy of the system. However, the numerical results exhibit non-monotonic behavior in certain regions. While some discrepancies exist between the simple theoretical model and the numerical simulations results, the overall trends are in good agreement. The dependence of the critical temperature on anisotropy was fitted in Fig. \ref{fig:theory_vs_MD} using Eq. \eqref{toy-temperature}. Although the toy model does not fully capture the non-monotonic intervals of this dependence, it still provides a close approximation to the overall trend. One possible explanation for this discrepancy is that the toy-model approximation is only valid for the low anisotropy, where the agreement is indeed quite good.

To further validate our findings, we computed the diffusion coefficients for the 14-ion system across a range of temperatures and anisotropies. In principle, the diffusion coefficients provide an independent means of estimating the melting transition, complementing the structural analysis. However, due to significant uncertainties in the computed values, extracting precise critical temperatures from the diffusion data remains challenging. This highlights the need for further refinement of numerical techniques or alternative methodologies to improve the accuracy of diffusion-based melting estimations.

Overall, our study provides valuable insight into the melting dynamics of confined 2D ion crystals. The agreement between theoretical and numerical approaches supports the robustness of our model, while the observed deviations emphasize the complexity of melting in finite systems.

\unskip

Future work should explore larger ion ensembles and refine theoretical approximations to improve quantitative accuracy. One key objective is to refine the computation of the diffusion coefficient by reducing statistical uncertainties, thus enabling a more precise determination of the critical temperature. This will involve increasing the number of independent trajectories and optimizing averaging techniques.

Another promising approach is the calculation of the free energy, which would provide a more fundamental thermodynamic perspective on the melting transition. Implementing free energy methods, such as thermodynamic integration, will allow for a direct comparison with structural and dynamic criteria for melting.

Yet another alternative method for determining the $T_c$ involves analyzing the velocity autocorrelation function (VACF)\cite{stanly2011}. This approach provides insight into the dynamical properties of the system, particularly the transition from localized oscillatory motion in the crystalline phase to the diffusive motion in the liquid phase. By examining how the VACF evolves with temperature and anisotropy, it will be possible to establish an independent criterion for the melting transition, complementing both structural and diffusion-based analyses.

These future directions aim to enhance the accuracy and robustness of critical temperature estimation in 2D trapped ion crystals, providing deeper insights into phase transitions in confined Coulomb systems.

%%%%%%%%%%%%%%%%%%%%%%%%%%%%%%%%%%%%%%%%%%
\section{Conclusions}

We have explored the structural transitions and melting behavior of two-dimensional ion crystals in an RF trap, combining theoretical analysis, molecular dynamics simulations, and experimental insights. Our results show that the melting transition strongly depends on ion number and trap anisotropy, with larger systems exhibiting lower critical temperatures. We identified that melting occurs through a ring-by-ring mechanism, with outer ions losing structural order first.

Despite some deviations, our theoretical model provides a reasonable approximation of the melting behavior, capturing key trends in the phase diagram. Additionally, we compared different approaches for determining the melting transition, showing that diffusion-based methods, while useful, introduce larger uncertainties than probability-based calculations.

These findings contribute to the broader understanding of phase transitions in confined Coulomb systems and offer insights for optimizing ion-crystal-based platforms for quantum simulations and many-body physics. Future work will refine the theoretical model and extend numerical analyses to improve accuracy in determining critical temperatures.

\bibliography{bib}
\end{document}